# REGULARIZATION OF INVERS PROBLEM FOR M-ARY CHANNEL

N. A. Filimonova

Novosibirsk, Russia

**Abstract.** The problem of computation of parameters of *m*-ary channel is considered. It is demonstrated that although the problem is ill-posed, it is possible "turning" of the parameters of the system and transform the problem to well-posed one.

**Statement of the problem.** We analyze the well-known formula for probability of correct identification if orthogonal signal in *m*-ary channel, which has the form

$$q(\delta, P_s, P_n, B, m) = \frac{1}{\sqrt{2\pi}} \int_{-\infty}^{\infty} \exp[-(z-(1-\delta)g\sqrt{B})^2/2] \, F(z)^{m-1} dz, \qquad (1)$$

where [1]

- $F(z) = \frac{1}{\sqrt{2\pi}} \int_{-\infty}^{z} \exp(-t^2/2) dt$,
- $g^2 = P_s / P_n$ is «signal to noise» ratio ($P_s$ and $P_n$ are averaged powers of signal and noise),
- $B$ is «base»a of signal (duration of signal multiplied by the specter width),
- *m* is dimension of signal,
- *d* is cancel interval thickness,
- $\delta = d/I$ is relatively cancel interval thickness.

The problem under consideration is formulated as follows: one has to determine a parameter of *M*-ary channel, if probability $q = q^*$ is known (from experiment, analysis of statistics etc.). In other words, one has to solve equation $q(\delta, P_s, P_n, B, m) = q^*$ with respect to one of the parameters $\delta, P_s, P_n, B, m$.

Observing formula (1), we find that the function $q(\delta, P_s, P_n, B, m)$ of the arguments $\delta, P_s, P_n, B$ depends, in fact, on the variable (invariant)

$$x = (1-\delta)g\sqrt{B}. \qquad (2)$$

and has the form

$$q(\delta, P_s, P_n, B, m) = Q_m(x). \qquad (3)$$

By virtue of (3) and (4), the inverse problem can be written in the terms of the invariant $x$

$$Q_m(x) = q^*. \qquad (4)$$

**Results of numerical analysis of formula (1).** Plots of function (3) of the argument $x$ were drown (using Mathcad software) for various *m*. The plots are shown at Fig.1.

It is seen from Fig.1 that the problem (4) is unstable with respect to the right-hand side $q^*$ for $q^* \approx 1$ and for small $q^*$ when *m* is large (100 and greater). At the same time, we see from Fig.2 that for every *m* there exists interval $[a_m, b_m]$ where the problem (4) is well-posed. The number $a_m = 0$ for small *m*, and $3 \leq b_m \leq 5$.

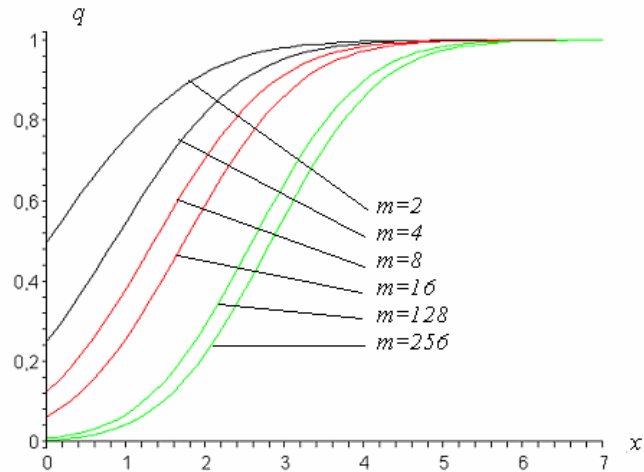

**Fig.1.** The plots of the function $q = Q_m(x)$ for various $m$

**Regularization of the problem by "turning" of parameters of the channel.** The original problem (3) is solved with respect to one of the variables $\delta, P_s, P_n, B$, not with respect to the invariant $x$. If we know interval of possible values of the unknown variable, we can use the remaining variables and give them values such that invariant $x \in [a_m, b_m]$. In this case the problem (4) can be solved with high accuracy with respect to the invariant $x$ and then the unknown variable can be computed.

Thus, on the set

$$a_m \leq (1-\delta)g\sqrt{B} \leq b_m. \qquad (5)$$

the problem of determining of a parameter of $m$-ary channel is well-posed.

In the technical terms our results means the following. In the general, the problem of determining of a parameter of $m$-ary channel is ill-posed. But it is well-posed if one tunes devices in the appropriate way. The condition (5) is the condition of the appropriate "tuning" of devices forming $m$-ary channel. The term "turning" in this paper corresponds to turning of real equipments.